 \documentclass[jip]{degruyter-journal-a}      

\title{Geo-information system of spread of tuberculosis based on inversion and prediction}
\headlinetitle{GIS of tuberculosis based on inversion and prediction}

\lastnameone{Kabanikhin}
\firstnameone{Sergey}
\nameshortone{S.~Kabanikhin}
\addressone{Institute of Computational Mathematics and Mathematical Geophysics SB RAS, Novosibirsk State University}
\countryone{Russia}
\emailone{kabanikhin@sscc.ru}

\lastnametwo{Krivorotko}
\firstnametwo{Olga}
\nameshorttwo{O.~Krivorotko}
\addresstwo{Novosibirsk State University, Institute of Computational Mathematics and Mathematical Geophysics SB RAS}
\countrytwo{Russia}
\emailtwo{krivorotko.olya@mail.ru}

\lastnamethree{Takuadina}
\firstnamethree{Aliya}
\nameshortthree{A.~Takuadina}
\addressthree{L.N. Gumilyov Eurasian National University}
\countrythree{Kazakhstan}
\emailthree{alyoka.01@mail.ru}

\lastnamefour{Andornaya}
\firstnamefour{Darya}
\nameshortfour{D.~Andornaya}
\addressfour{Novosibirsk State University}
\countryfour{Russia}
\emailfour{ermolenko.dasha@mail.ru}

\lastnamefive{Zhang}
\firstnamefive{Shuhua}
\nameshortfive{S.~Zhang}
\addressfive{Tianjin University of Finance and Economics}
\countryfive{China}
\emailfive{shuhua55@126.com}

\headlineauthor{S. Kabanikhin, O. Krivorotko, A. Takuadina et al.}

\abstract{The monitoring, analysis and prediction of epidemic spread in the region require the construction of mathematical model, big data processing and visualization because the amount of population and the size of the region could be huge. One of the important steps is refinement of mathematical model, i.e. determination of initial data and coefficients of system of differential equations which describe the epidemiology processes. We analyze numerical method for solving inverse problem of epidemiology based on genetic algorithm and traditional optimization ideas. Numerical results are applied to analysis and prediction of epidemic situation in regions of Russian Federation, Republic of Kazakhstan and People's Republic of China. Due to a great amount of data we use a special Geo-information system for visualization of epidemic process, i.e. a special software named Digital Earth.}



\researchsupported{This work is supported by the Russian Science Foundation (project No.~18-71-10044).}

\acknowledgments{Authors are grateful to Igor Marinin for the help in construction of prediction map with Digital Earth.}

\begin{document}
\section*{Introduction}
A mathematical model of the spread of an infectious disease in a population describes the transmission of a pathogen depending on the nature of contacts among infectious and susceptible people, the latent period of infection, the duration of infection, the degree of acquired immunity after infection, etc. After all these factors are taken into account in the model, we can do forecasts regarding the number of people who are expected to be infected during the epidemic, the duration of the epidemic, and the maximum incidence rate. We describe the stages of solving the problems of refining the parameters of mathematical models for forecasting the tuberculosis epidemic using the example of the Karaganda region, as well as constructing three-dimensional forecast maps.

In paper \cite{1} we introduced a combined numerical algorithm for reconstructing a mathematical model to describe tuberculosis transmission. In the present paper we propose a stochastic approach for parameter reconstruction in mathematical model for tuberculosis transmission with control programs for Karaganda region. 
The article is organized as follows. In Section 1 we describe the mathematical model of epidemiology with the emergence of drug-resistant strains. Then we formulate the inverse problem (Section 2) and genetic algorithm for solving this problem (Section 3). In Section 4 we describe results of numerical calculations. In Section 5 we apply numerical results to Geo-information system and present tuberculosis maps.

\section {A mathematical model of the spread of tuberculosis with the emergence of drug-resistant strains}

For modelling the epidemiology processes, nonlinear systems of ordinary differential equations (ODE) describing the mass balance law are used. The coefficients of systems of ODE characterize the population and development of the disease. In order to refine the mathematical model for a specific population, a qualitative assessment of the model parameters or their combinations is necessary \cite{2}. The Cauchy problem for the mathematical modeling of dynamic processes is based on the mass conservation law:
\begin{equation}
\label{trivial}
\dot X  = F (X (t),\Theta),\ X (t_0)= X^0,\quad t \in (t_0,T),\ T> t_0.
\end{equation}
Here $X (t)= (X_1  (t),\ldots ,X_N  (t))^T$ is a vector of functions that can describe the number of carriers of infection, non-infectious patients, patients on treatment, etc., $ \Theta =(\Theta_1,\ldots,\Theta_M)^T\in F $ is the parameter vector characterizing the transmission rate of the infection, the probability of recovery or mortality from the disease, etc., $F: =\{p \in \mathbb{R}^M: p_m  \ge 0,m = 1,\ldots,M\}$ is the parameter space under consideration, $F (X (t),\Theta) = (F_1  (X (t),\Theta),...,F_N  (X (t),\Theta))^T$ is a vector function 
$F_n  (X (t),\Theta): C^2  (t_0,T) \to C^2  (t_0,T), \ n = 1,\ldots,N$, and $X^0  = (X_1^0  ,...,X_N^0) ^T$ is the vector of the initial data. 

A mathematical model of the spread of an infectious disease in a population describes the transmission of a pathogen depending on the nature of contacts among infectious and susceptible people, the latent period of infection, the duration of infection, the degree of acquired immunity after infection, etc. After all these factors are formulated in the model, we can do forecasts regarding the number of people who are expected to be infected during the epidemic, the duration of the epidemic, and the maximum incidence rate \cite{3}.

Consider the Cauchy problem for a mathematical model of the spread of tuberculosis in highly endemic regions, proposed by James Trauer et al. \cite{4,5}:
\begin{equation}
\label{Krug}
\begin{cases}
\frac{dS}{dt}=l\Pi +\varphi T+\varphi _m T_m-(\lambda_d+\lambda_
{dm}+\mu)S,\\
\frac{dL_A}{dt}=\lambda_d (S+L_B+L_{Bm} )-(\varepsilon+k+\mu) L_A,  \\ 
\frac{dL_{Am}}{dt}=\lambda_{dm} (S+L_B+L_{Bm} )-(\varepsilon+k+\mu) L_{Am}, \\
\frac{dL_B}{dt}=kL_A+\gamma I-(\lambda_d+\lambda_{dm}+\nu+\mu) L_B, \\
\frac{dL_{Bm}}{dt}=kL_{Am}+\gamma I_m-(\lambda_d+\lambda_{dm}+\nu+\mu) L_{Bm}, \\
\frac{dI}{dt}=\varepsilon L_A+\nu L_B+(1-\eta)\omega T-(\gamma+\delta+\mu_i )I, \\
\frac{dI_m}{dt}=\varepsilon L_{Am}+\nu L_{Bm}+\eta \omega T-(\gamma+\delta_m+\mu_i ) I_m, \\
\frac{dT}{dt}=\delta I-(\varphi+\omega+\mu_t )T, \\
\frac{dT_m}{dt}=\delta_m I_m-(\varphi_m+\omega+\mu_t ) T_m,  \\
S(t_0)=S_0, \ L_A (t_0)=L_{A_0}, \ L_{Am} (t_0)=L_{Am_0}, \\
L_B (t_0)=L_{B_0}, \  L_{Bm} (t_0)=L_{Bm_0},\ I(t_0)=I_0, \\
I_m (t_0)=I_{m_0},\ T(t_0)=T_0,\ T_m (t_0)=T_{m_0}.
\end{cases}
\end{equation}
Here $t\in (t_0,T)$ is a time variable measured in years and
\[ \lambda_d=\chi\beta\rho(I+o T)/N,  \  \lambda_{dm}=\chi\beta_m \rho(I_m+o T_m)/N.\]

In the Cauchy problem~(\ref{Krug}) the population is divided into vaccinated sensitive (usually children under 14) ($S$), latent carriers of the infection, having an MDR-TB strain (index $m$), with fast ($L_A, L_{Am}$) and slow ($L_B, L_{Bm}$) development active forms of the disease, infected patients undergoing treatment for the disease ($T, T_m$) and not being treated ($I, I_m$). $N = S + L_A + L_{Am} + L_B + L_{Bm} + I + I_m + T + T_m$ is the whole population. The description and average value of the model parameters~(\ref{Krug}) are given in Table~\ref{model_parameters}.

\begin{table}
\caption{Description and values of parameters of the mathematical model~(\ref{Krug}) for the Karaganda region.}
\label{model_parameters}
\begin{tabular}{|p{1cm}|p{5.5cm}|p{1.6cm}|p{2.5cm}|}
\hline
Symbol & Description & Units & Size\\ \hline

\hline
$\Pi$ & inflow of young people to model population & people/year & depends on population type\\
$N$ & total population size  & people & depends on population type\\
$\mu$ & the coefficient of mortality from
all causes except for tuberculosis & year & 0.016\\
$\varphi$ & tuberculosis treatment rate & people/year & 2\\
$\varphi_m$ & MDR-TB treatment rate & people/year & 0.5\\
$\varepsilon$ & disease early progression rate & year & 0.129\\
$k$ & transition rate to disease late progression & year & 0.821\\
$\gamma$ & spontaneous self-recovery rate & year & 0.63\\
$\nu$ & development rate of active disease
form under endogenous activation & year & 0.075\\
$\eta$ & probability of MDR-TB strain development during treatment & - & 0.035\\
$\omega$ & reinfection rate & people/year & 0.25\\
$\delta$ & detection rate of individuals with active TB form & people/year & 0.72\\
$\delta_m$ & detection rate of individuals with active MDR-TB form & people/year & 0.035\\
$\mu_i$ & tuberculosis mortality without treatment & year & 0.37\\
$\mu_t$ & tuberculosis mortality during treatment & year & $0.5 \mu_i$\\
$\beta$ & contagiousness parameter & - & depends on population type\\
$\beta_m$ & contagiousness parameter for MDR-TB & - & 0.7 $\beta$\\
$\chi$ & partial immunity parameter & - & 0.49\\
$\rho$ & infection fraction & - & 0.7\\
$o$ & transmissibility parameter of individuals during treatment & - & 0.6\\
$l$ & BCG vaccination rate & - & 0.65\\\hline
\end{tabular}
\end{table}

Since in Russian Federation and Republic of Kazakhstan according to the legislation, vaccination of BCG infants (Bacillus Calmette - Guerin) is mandatory for citizens whose incidence exceeds 80 cases per 100,000 people, and the majority of the population was vaccinated at preschool age. Therefore, in model~(\ref{Krug}) there is no division of the population into sensitive unvaccinated and vaccinated~\cite{1}.

\section {Inverse problem}
Model parameters can be estimated (or sometimes uniquely determined) using some additional information about biological processes (number of infected or treated patients). The task of determining biological parameters using such additional information is called the inverse problem \cite{3} which in the general case is ill-posed.
The inverse problem for direct problem~(\ref{Krug}) consists in determination of the vector of model parameters and the initial data $q = (\Theta, X ^ 0) ^ T$ for a given logistic function $F$ as well as some additional information about a part of the components of the vector $X (t; q)$:
\begin{equation}
\label{thry}
 X_n  (t_k; q)= \Phi_n  (t_k ),\; t_k \in (t_0,T),\, k = 1,\ldots ,K_n, \ n \in I\subseteq I_N. 
\end{equation}

Here $I_N:=\{1,..., N\}$, $\Phi_n (t_k): = \Phi _ n ^ {(k)} = (\Phi _ n ^ {(1)}, ..., \Phi_ n ^ {(K_n)}) ^ T$ is the data vector of the inverse problem of dimension $K = K_1 + \ldots+ K_N.$ 
The vector $\Phi = (\Phi_1 ^ {(1)}, ..., \Phi_1 ^ {(K_1)}, ..., \Phi_N ^ {(1)}, ..., \Phi_N ^ {(K_N)}) ^ T$ is determined by statistic data at time $t_k, k = 1, ..., K_n.$ Suppose, that measured components of the vector $X (t; q)$ are equal to $H\leq N$.
We define an operator for the inverse problem~(\ref{Krug})-(\ref{thry}) as follows: $A: q \in \mathbb{R} ^ {M + H} \rightarrow \Phi \in \mathbb{R} ^ K.$ Thus, the inverse problem~(\ref{Krug})-(\ref{thry}) can be written in the operator form $A (q) = \Phi$. We reduce the inverse problem to the problem of minimizing the following objective functional: 
\begin{equation}
\label{fohr}
 J (q)= \|A (q)-\Phi\|^2  =  \frac {T-t_0}{H\cdot K} \sum_{n=1}^H \sum_{k=1}^K \bigl| X_n  (t_k; q)-\Phi_n^{(k)} \bigr|^2 . 
\end{equation}  
The problem of minimizing functional~(\ref{fohr}) is solved using the genetic algorithm.

\subsection{Identifiability analysis}
The solution $q$ of the inverse problem~(\ref{Krug})-(\ref{thry}) could be non-unique or/and unstable to the data noise. It is necessary to conduct the identifiability analysis before construct the regularization algorithm for solving of an inverse problem. It helps to answer on questions:\\
-- How to verify whether the model parameters are identifiable based on the measurements of output variables or their functions when the ODE model does not have a closed-form solution?\\
-- If not all model parameters are identifiable, are a subset of parameters identifiable?\\
-- How many measurements, at which time points, are necessary to identify the identifiable parameters?

The review on identifiability including structural, practical identifiability as well as sensitivity-based analysis for nonlinear ODE is given in paper~\cite{11}. The structural identifiability depends on a model structure $F(X(t),\Theta)$ and uses linear and differential algebra approaches that based on exclusion of $X(t)$ and obtaining the relationship on $\Phi$~\cite{KSI_Voronov_identifiability2016}. The practical identifiability learns the deviation of noise level in measured data~(\ref{thry}) on the volume of confidence interval of solution $q$~\cite{Banks2014}. The sensitivity-based identifiability analysis is similar to the structural analysis approach in the sense that both approaches do not require actual experimental data, and both approaches assume that measurements are precise without error. On the other hand, the sensitivity-based method is similar to the practical analysis approach in the sense that both methods require pre-specified parameter values, and both need to know the number and locations of measurement time points~\cite{KOI_Latushenko_identifiability2018}.

Sensitivity-based identifiability analysis is used to assess the identifiability of the unknown model parameters $\Theta$ for the system of ODEs~(\ref{trivial}),~(\ref{thry}) and is based on the investigation of sensitivity matrix for the vector of parameters $\Theta^*$ with coefficients:
\[s_{ij}(t) = \frac{\partial \Phi_i(t;\Theta^*)}{\partial \Theta_j}.\]

To calculate the sensitivity matrix $S\in \mathbb{R}^{H\cdot K\times M}$ we solve the Cauchy problem for the sensitivity function $s_{\Theta_i} = \frac{\partial X}{\partial\Theta_i}$, $i=1,\ldots,M$:
\begin{eqnarray*}
    \left\{\begin{array}{ll}
        \dot{s}_{\Theta_i}(t) & = \frac{\partial F}{\partial X}(X(t;\Theta),\Theta)s_{\Theta_i} + \frac{\partial F}{\partial \Theta_i}(X(t;\Theta),\Theta),\\
        s_{\Theta_i}(t_0) & = 0.
    \end{array}\right.
\end{eqnarray*}

For the investigation of sensitivity matrix $S$ we apply the eigenvalue method that was proposed in paper~\cite{10}. This method is based on properties of eigenvalues and eigenvectors of the matrix with components:
\[H_{ml} = \sum_{i,k}\frac{\partial \Phi_i(t_k;\Theta^*)}{\partial \Theta_l}\cdot \frac{\partial\Phi_i(t_k;\Theta^*)}{\partial q_m}=(S^T S)_{ml}.\]

\begin{figure}[!ht]
\center{\includegraphics[scale=0.6]{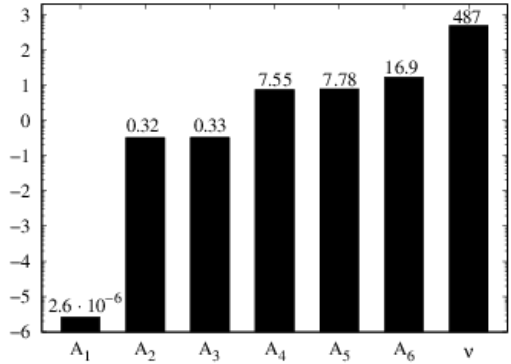}}
\caption{The minimum eigenvalues for the matrix $H$ on a logarithmic scale. Here $A_1$ is a set of all parameters $q$, $A_2=A_1\setminus\{o\}$, $A_3=A_2\setminus\{\varepsilon \}$, $A_4=A_3\setminus\{\delta_m\}$, $A_5=A_4\setminus\{\rho \chi\}$, $A_6 = A_5\setminus\{\delta\}$.}
\label{Identifiability_eigenvalue}
\end{figure}

We will determine only those parameters in model~(\ref{Krug}) that directly affect the spread of the epidemic in a particular region $\Theta = (\varepsilon, k, \nu, \delta, \delta_m, \rho \chi, o)\in\mathbb{R}^7.$ The most sensitive parameter is determined at each iteration of the algorithm by analyzing of the minimum eigenvalue of the matrix $H$. At the next iteration the most sensitive parameter is excluded from consideration, and the matrix $H$ is reorganized for a new set of parameters. The set of parameters considered at each iteration of the algorithm is shown under each column in Fig.~\ref{Identifiability_eigenvalue}. The sequence of coefficients of the model~(\ref{Krug}) from the least to the most sensitive to noise in measurements~(\ref{thry}) is follows:
\[\nu, \delta, \rho\chi, \delta_m, \varepsilon, o, k.\]
Note, that for coefficient $k$ that characterizes transition rate to disease  late progression the minimum eigenvalue of $H$ is equal to $2.6\cdot 10^{-6}$. In numerical calculation we should pay attention to stability of this parameter to the noise level in measurements~(\ref{thry}).

\section {Methods for solving the inverse problem}
The most effective methods for solving optimization problems include stochastic methods (genetic algorithms, Monte Carlo method, simulated annealing, etc.) and deterministic methods (Levenberg-Marquardt method, Landweber iteration, conjugate gradients, Nelder-Mead, etc.).
\begin{itemize}
\item Deterministic methods are used only for single-purpose optimization problems, have differentiability requirements for the objective function, require a good initial approximation, work unstably on a large number of parameters (that is, they do not guarantee finding the global minimum of the problem).
\item Stochastic methods are more widespread than deterministic ones since they do not require the smoothness of target functional, and they have the ability to circumvent local extrema and are used to solve multioptimization problems. Among stochastic methods, the most common are methods based on a simpler task. For example, genetic algorithms make it possible to reduce the solution of a complex problem to the solution of the problem of developing a population of living creatures with certain properties.
\end{itemize}

\subsection{Genetic algorithm}
Genetic algorithm searches for the optimal solution, while simultaneously analyzing the many current sets of parameters, called individuals, that have evolved over many generations in accordance with the prescribed rules. The need for a large number of calculations of the target functional is compensated by the possibility of parallel calculation. The genetic algorithm has been successfully used to solve the problem of optimization design \cite{10}. The block diagram (Fig.~\ref{The block}) and structure of the algorithm are given below.

\begin{figure}[!ht]
\center{\includegraphics[width=1\linewidth]{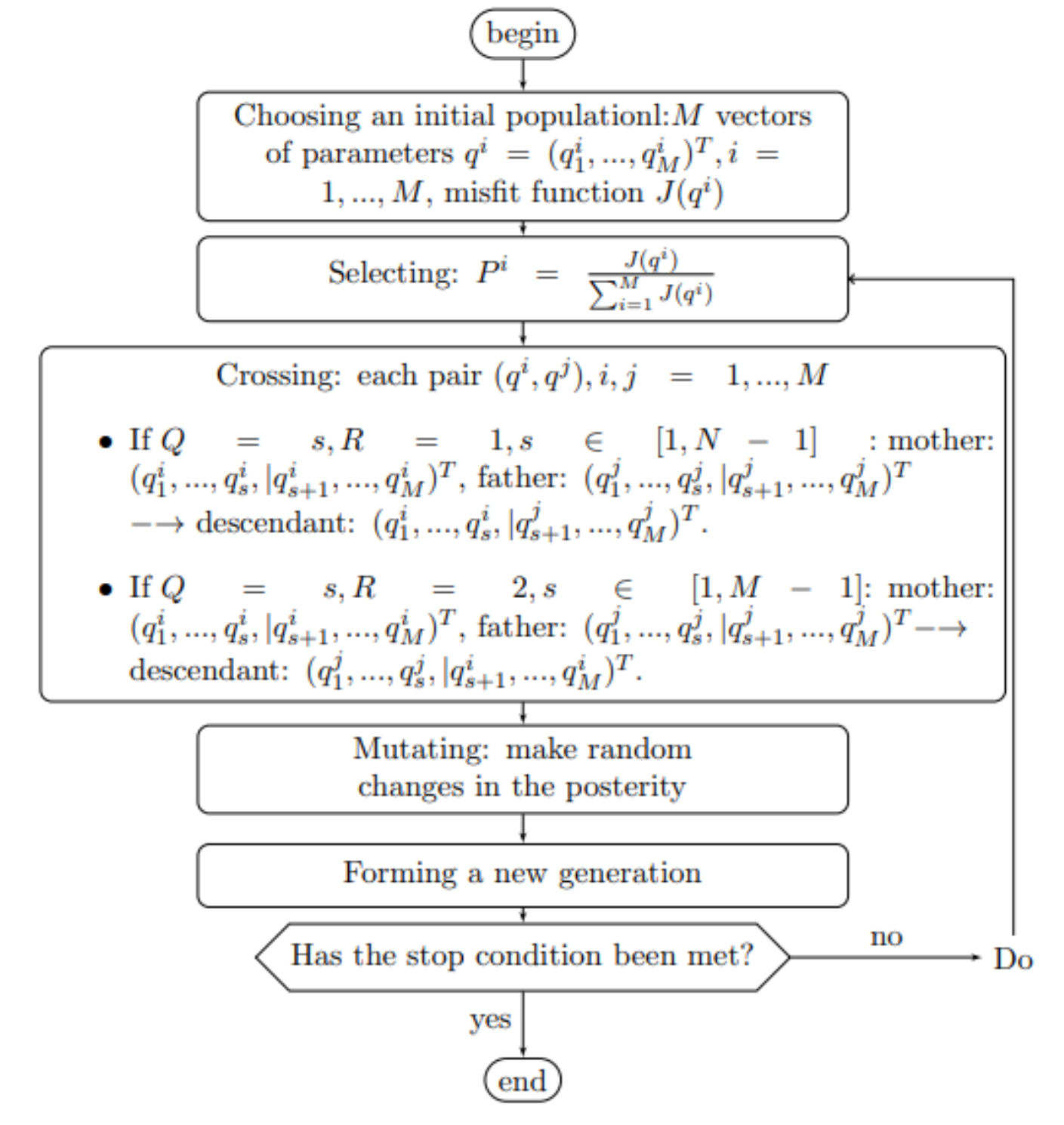}}
\caption{The block diagram of the genetic algorithm}
\label{The block}
\end{figure}

At the stage of Mutation:
\begin{itemize}
\item choose a whole number $A$ of descendants to which the mutation will be applied. Here $A$ is a random integer from 1 to $M$; 
\item Then choose whole number $B_i \in (1, M), i = 1, \ldots , A$. Here $B_i$, are the item numbers of descendants that will mutate; 
\item For each mutating descendant $B_i$ , choose a random volume $C_{B_i} , i = 1, \ldots, A $, of mutating elements. Here $C_{B_i}$ is a random integer from 1 to $M$ for each $i$; 
\item Then choose random integers $D_k \in (1, M), k = 1, \ldots , C_{B_i} , i = 1, \ldots, A, $ which characterize the item number of mutating elements and replace each mutating element by a new random value from the allowable period. 
\end{itemize}

Then сhoose the fittest member among the parents and descendants, i.e. select the member that has the lowest value of the misfit function $J(q^i)$. Also, choose a few "lucky ones" of the generation members that badly minimize the functional. They will bring diversity to the subsequent generations.

If at least one of the conditions is satisfied, the resulting population is found:
\begin{itemize}
\item $J(q^i ) < \Delta$. Here $J(q^i )$ is the lowest value of the misfit function in the population, and $\Delta$ is a predetermined number. In our paper, as the value of $\Delta$ we take 0.0001.
\item The smallest value of the misfit function in the population changes by less than $10^{-8}$ within 500 consecutive iterations.
\end{itemize}

Choose from the population a vector with the lowest value of the misfit function. If not, go to step 2.

If the genetic algorithm is stuck in a local minimum, the step of mutation will help to get out of it. As practice shows, we may find a global minimum using the genetic algorithm for the optimization problem. This is also confirmed by the statistical convergence theorem \cite{3}, which says that if in a population evolving with a given mutation scheme, the probability that the genetic algorithm will find the optimal solution asymptotically goes to one.

\section {Numerical calculations}
The coefficients of the Cauchy problem~(\ref{Krug}) is refined using statistics of Karaganda region of type (\ref{thry}) for 15 years using the genetic algorithm for minimization of misfit function~(\ref{fohr}). In subsection~\ref{iniset} the initial sets of numerical simulation are introduced. The synthetic noisy data of inverse problem~(\ref{Krug})-(\ref{thry}) are described in ~\ref{synthetic_data}. The results of numerical experiment are presented in subsection~\ref{num_experiment}.

\subsection{Initial sets}\label{iniset}
As a model example we consider the population of the Karaganda region of the Republic of Kazakhstan from 2000 to 2014 (15 years). The initial data for $t_0=2000$, taken for modeling (in thousand people):
\begin{eqnarray*}
\begin{array}{cc}
S(t_0 )=305.5,\ L_A (t_0 )=454.99, \ L_{Am} (t_0 )=71.09, \ L_B (t_0 )=457.46, \\ L_{B_m} (t_0 )=72.61, \ I(t_0 )=1.76, \ I_m (t_0 )=0.25, \  T(t_0 )=4.16, \  T_m (t_0 )=0.98.
\end{array}
\end{eqnarray*}

The population growth is $\Pi$ = 15.32 thousand people, the total number of people in the population is $N$ = 1368.8 thousand people, the mortality rate from all causes except tuberculosis is $\mu = 0.016$, and the contagiousness coefficient is $\beta$ = 0.025. Also note that the coefficients $\rho$ and $\chi$ in the model are always in the products. We will determine only those parameters that directly affect the spread of the epidemic in a particular region $q =\Theta = (\varepsilon, k, \nu, \delta, \delta_m, \rho \chi, o).$ The initial parameter vector, $q _ {(0)} \in\mathbb{R}^7$ is chosen arbitrarily in the interval $q _{{(0)}_i}\in [0,1] (i = 1,\ldots, 7)$. Therefore, we generate $q _{{(0)}_ i}$ as a quantity uniformly distributed in the interval $[0,1]$.

\subsection{Synthetic data}\label{synthetic_data}
To generate synthetic data, we select the vector of exact parameters
$q = (\varepsilon, k, \nu, \delta, $ $\delta_m, \rho \chi, o)$ taken from Table~\ref{model_parameters}. Solving the direct Cauchy problem~(\ref{Krug}) with parameter $q$ by the fourth-order Runge Kutta method approximations we obtain synthetic data $\Phi^k = (S_k, T_k, T_{m _k})$, distributed uniformly over the interval (0, T), the measurements of which occur once a year for fifteen years. Add to the data additive Gaussian noise with a level of 10  \% according to the formula:
\[\Phi_k = \Phi_k + 0.1 *\epsilon* \Phi_k.\] 
Here $\epsilon $  simulates a uniformly distributed Gaussian random variable. Data for three measured functions $\Phi^k= (S_k,T_k,T_{m_k})$ of the form~(\ref{thry}) with 10\% Gaussian noise are presented in Figure~\ref{fig:S, T and T_m}.
\begin{figure}[!ht]  
\centering{
\includegraphics[width=0.22\linewidth, angle=-90]{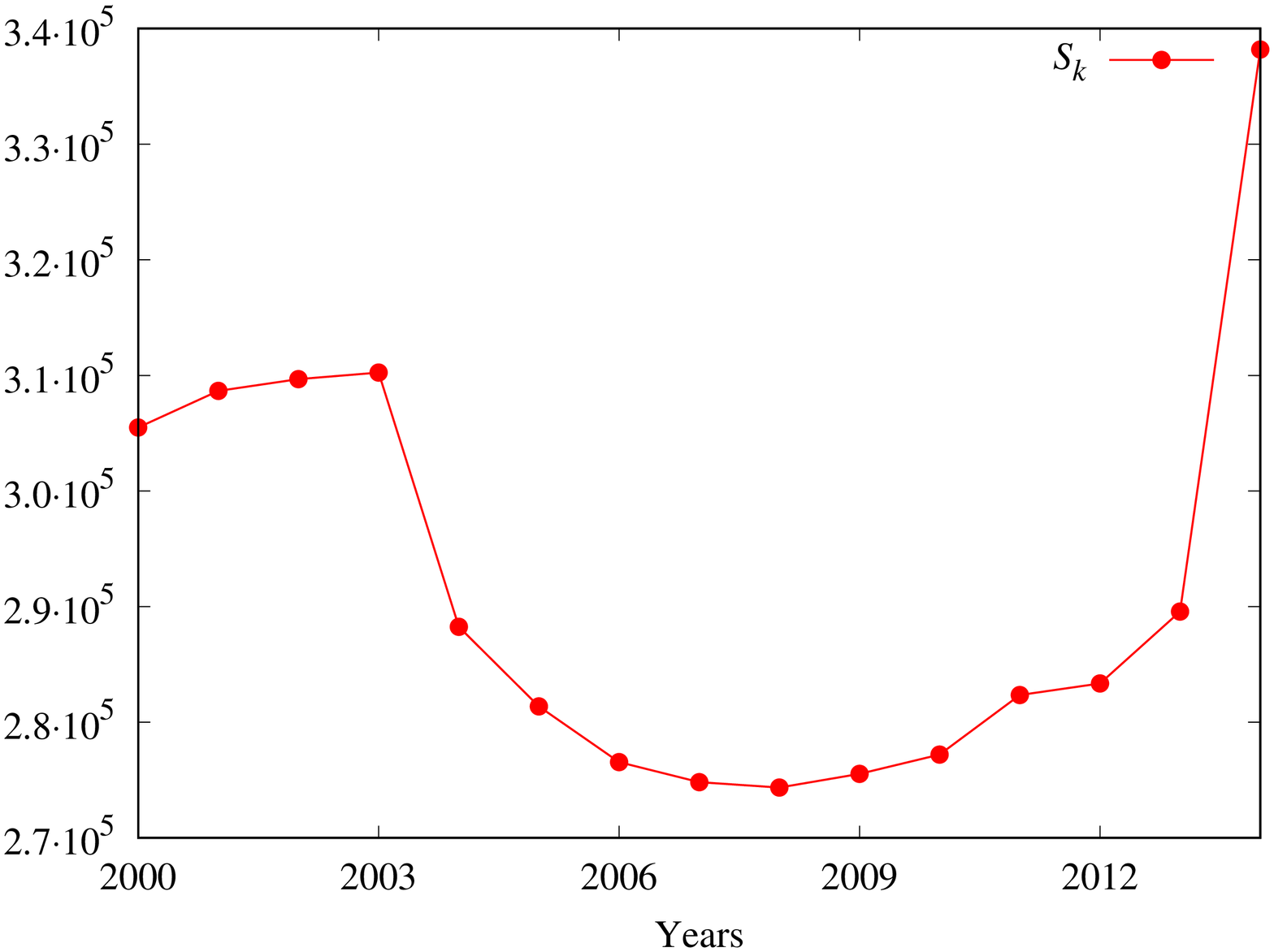}}  
\hspace{1ex}
{
\includegraphics[width=0.2\linewidth, angle=-90]{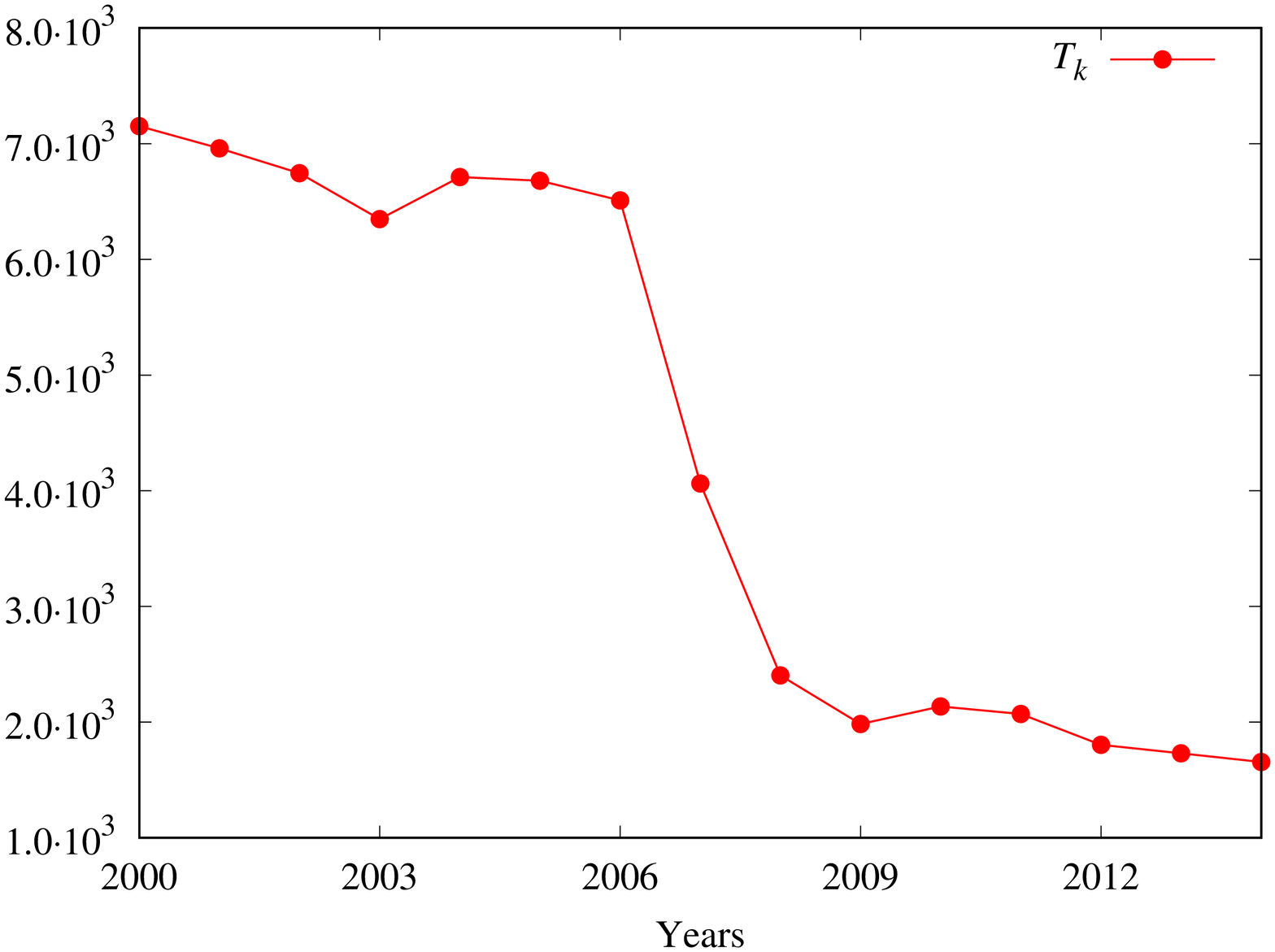}}
\hspace{1ex}
{ \includegraphics[width=0.2\linewidth, angle=-90]{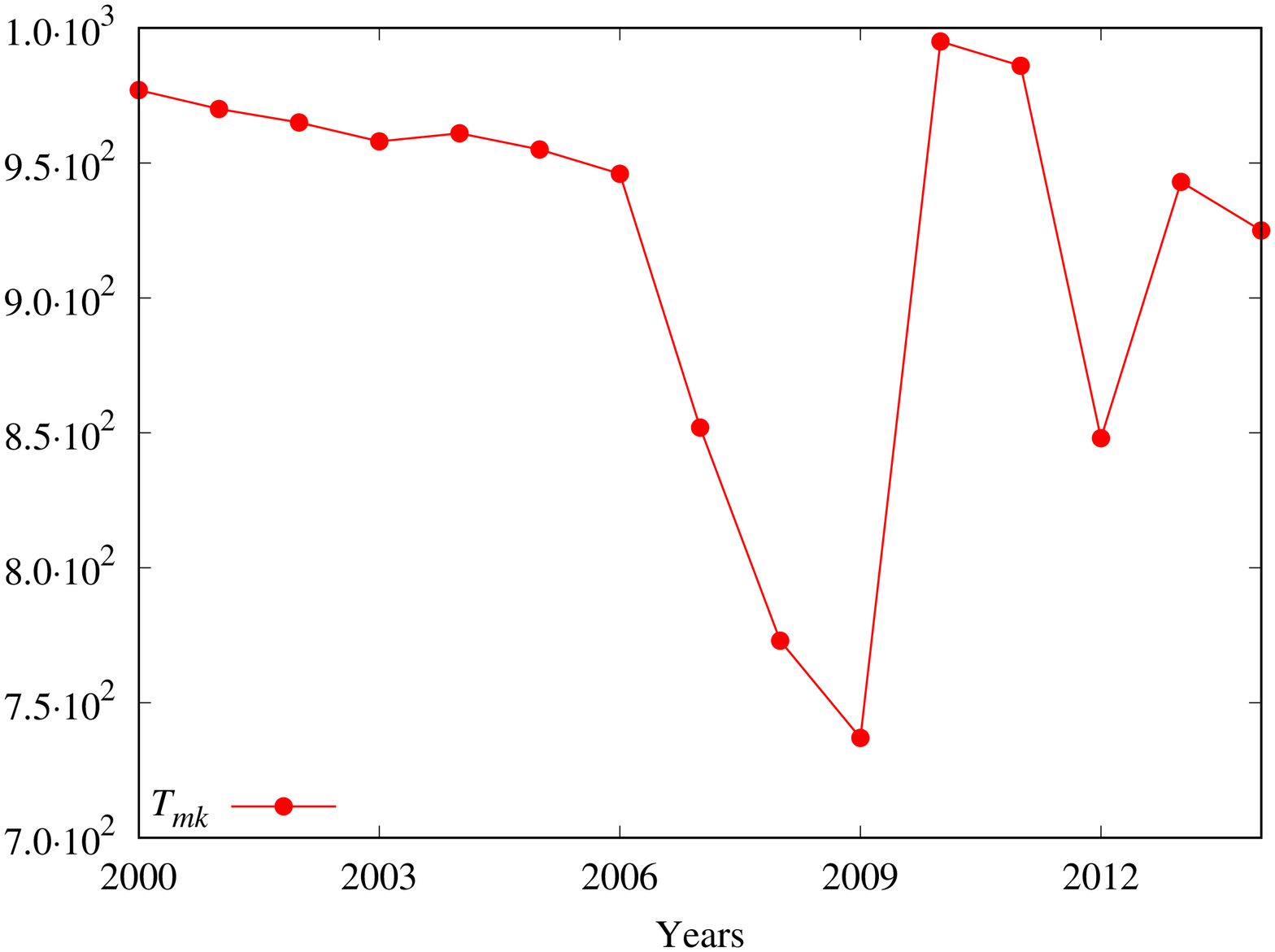}}  
\caption{Measurements of the inverse problem~(\ref{Krug})-(\ref{thry}) of vaccinated sensitive individuals $S_k$ (left), infected individuals under treatment for $t_k$ disease and having MDR TB strain $T_{m_k}$ from 2000 to 2014, $k=0,\ldots , 14.$} \label{fig:S, T and T_m} \end{figure}

\subsection{Numerical experiment}\label{num_experiment}
\begin{figure}[!ht]
\center{\includegraphics[scale=0.6]{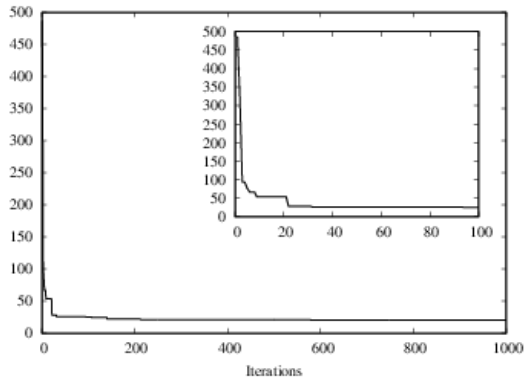}}
\caption{The value of the target functional~(\ref{fohr}) depending on the iterations of the genetic algorithm.}
\label{Functional}
\end{figure}

Solving the inverse problem in the case of perturbed data using the genetic algorithm, we have obtained the following results. Figure~\ref{Functional} shows a graph of the decrease of the target functional~(\ref{fohr}) depending on the number of iterations. The value of the functional in the execution of 7281 iterations is 24365.

\begin{figure}[!ht]
\begin{minipage}[h]{0.47\linewidth}
\center{\includegraphics[width=1\linewidth]{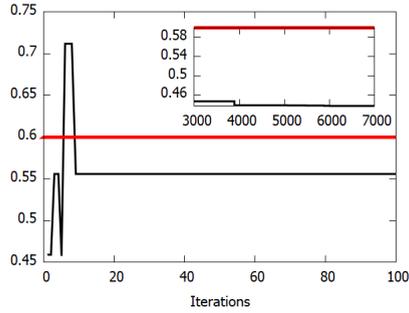}} g) Speed of treatment of infection $o$.\\
\end{minipage}
\caption{Dependence of parameter recovery $q = (\varepsilon, k,\nu, \delta, \delta_m, \rho \chi, o)$ on the number of iterations of the genetic algorithm.}
\label{ris:experimentalcorrelationsignals}
\end{figure}

Figures~\ref{ris:experimentalcorrelationsignals}a-g show graphs of the reconstructed parameters depending on the iterations of the genetic algorithm (black lines) and the exact value of each parameter (red lines). Quantitative characteristics of the reconstruction of the parameter vector $q \in \mathbb{R}^7$ are given in Table~\ref{tab_parameters}. Note that not all parameters are recovered with acceptable accuracy due to sensitivity to measurement errors~\cite{11}. However, the total relative error (see Table~\ref{tab_parameters}) does not exceed 15.5\%.

\begin{figure}[!ht]
\begin{minipage}[h]{0.47\linewidth}
\center{\includegraphics[width=1\linewidth]{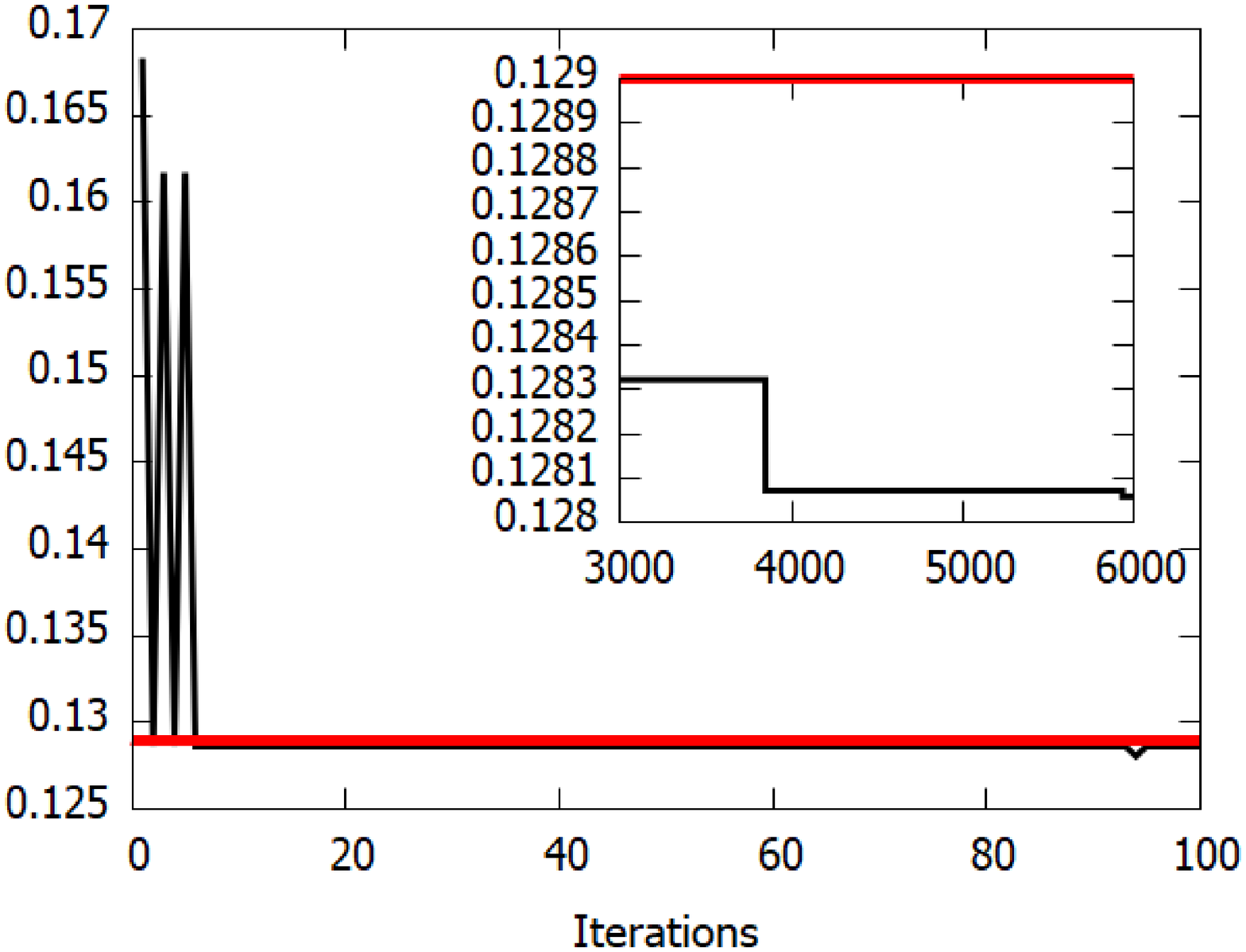}} a) The rate of early progression of the disease $\varepsilon. $\\
\end{minipage}
\hfill
\begin{minipage}[h]{0.47\linewidth}
\center{\includegraphics[width=1\linewidth]{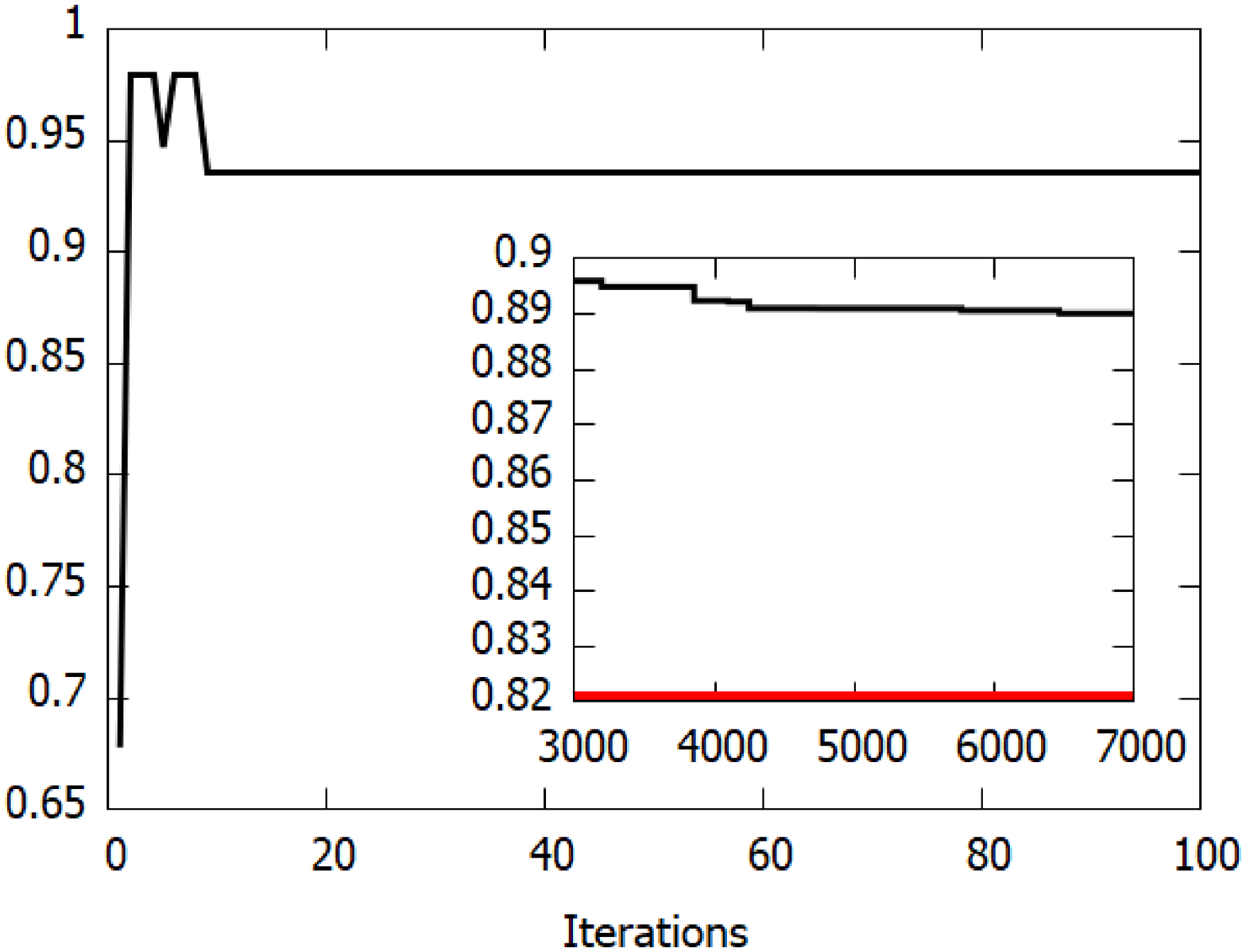}} \\b) The rate of transition to late disease progression $k$.
\end{minipage}
\vfill
\begin{minipage}[h]{0.47\linewidth}
\center{\includegraphics[width=1\linewidth]{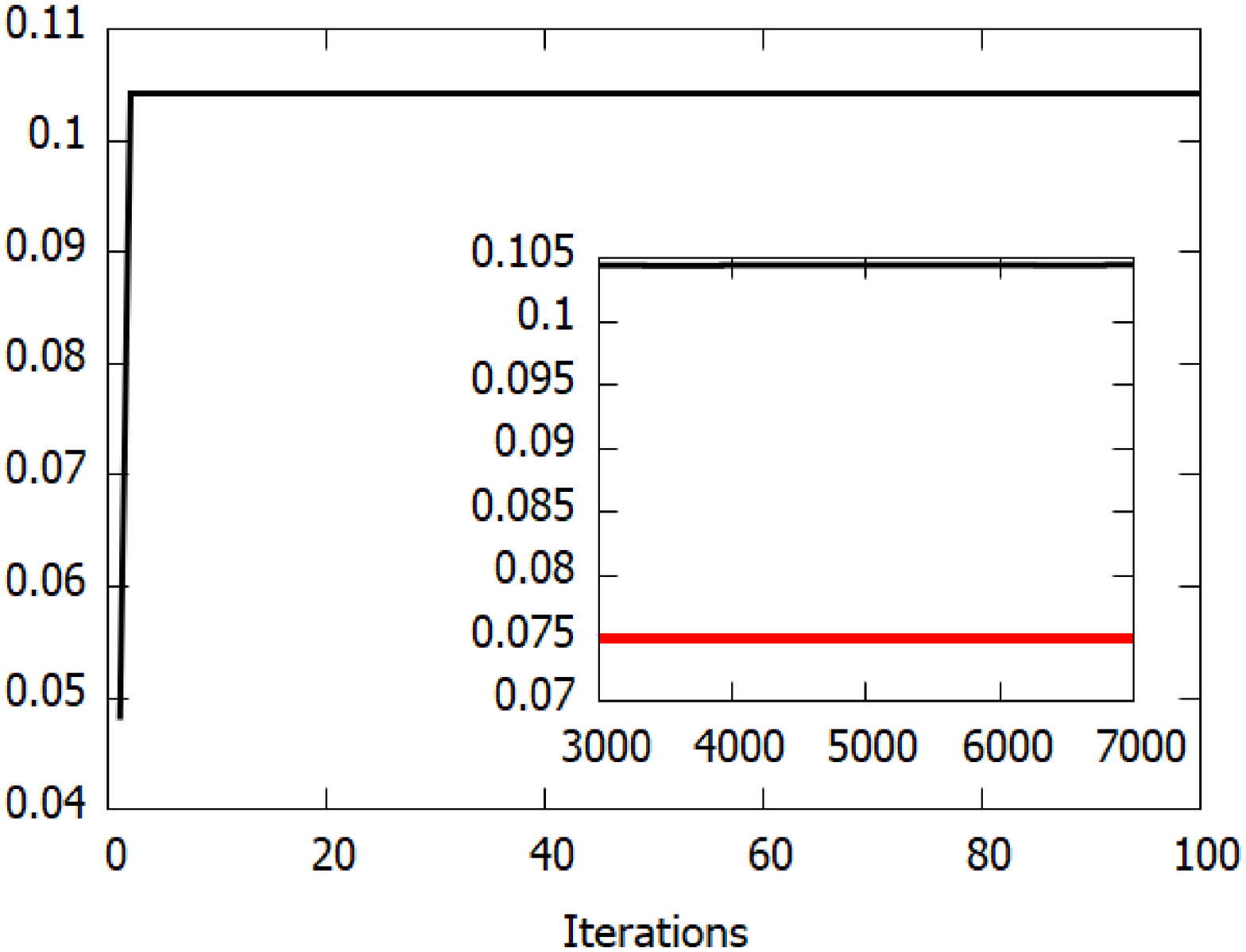}} c) The rate of development of the active form of the disease with endogenous activation $\nu.$\\
\end{minipage}
\hfill
\begin{minipage}[h]{0.47\linewidth}
\center{\includegraphics[width=1\linewidth]{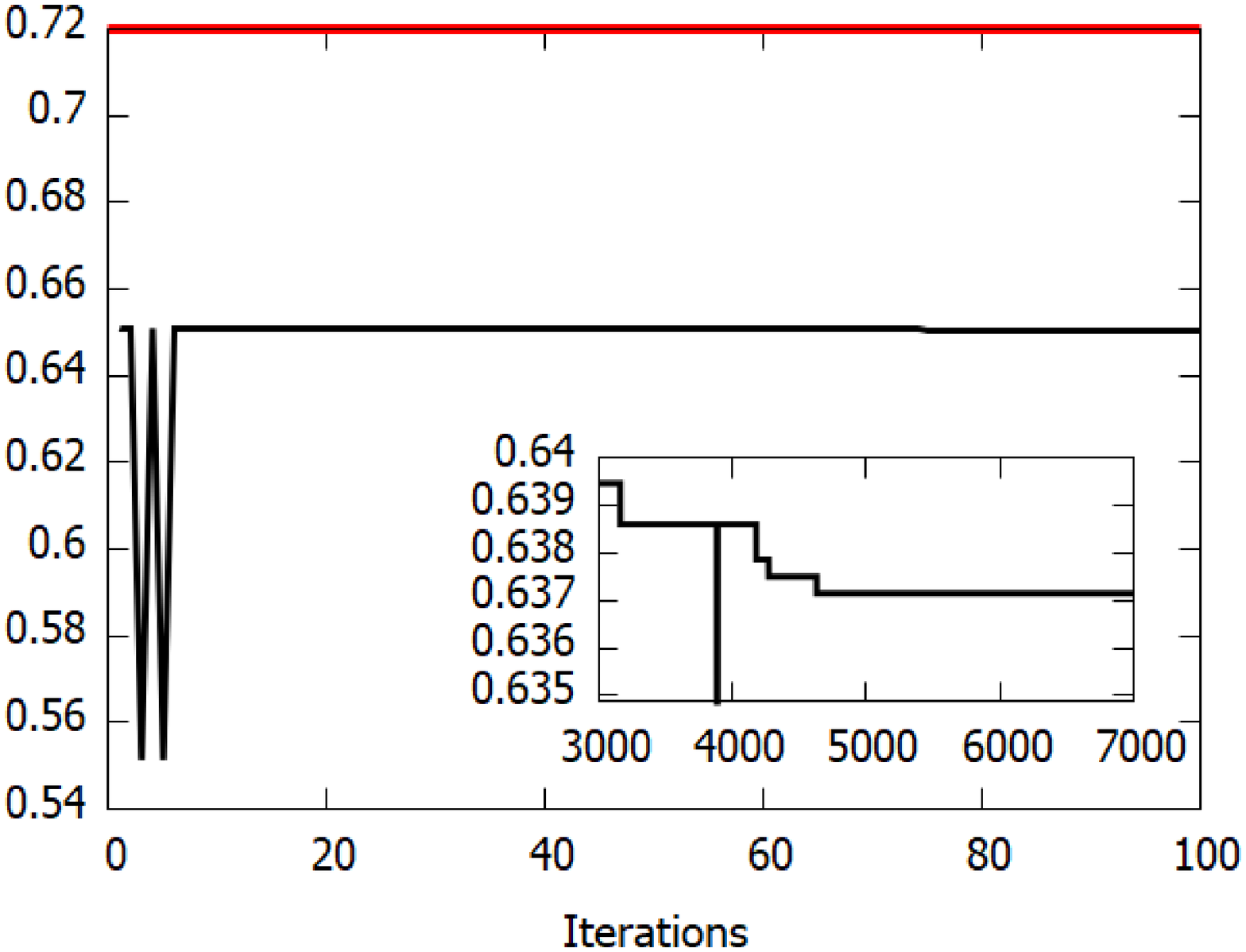}} \\d) Frequency of detection of persons with active tuberculosis $\delta$.
\end{minipage}
\vfill
\begin{minipage}[h]{0.47\linewidth}
\center{\includegraphics[width=1\linewidth]{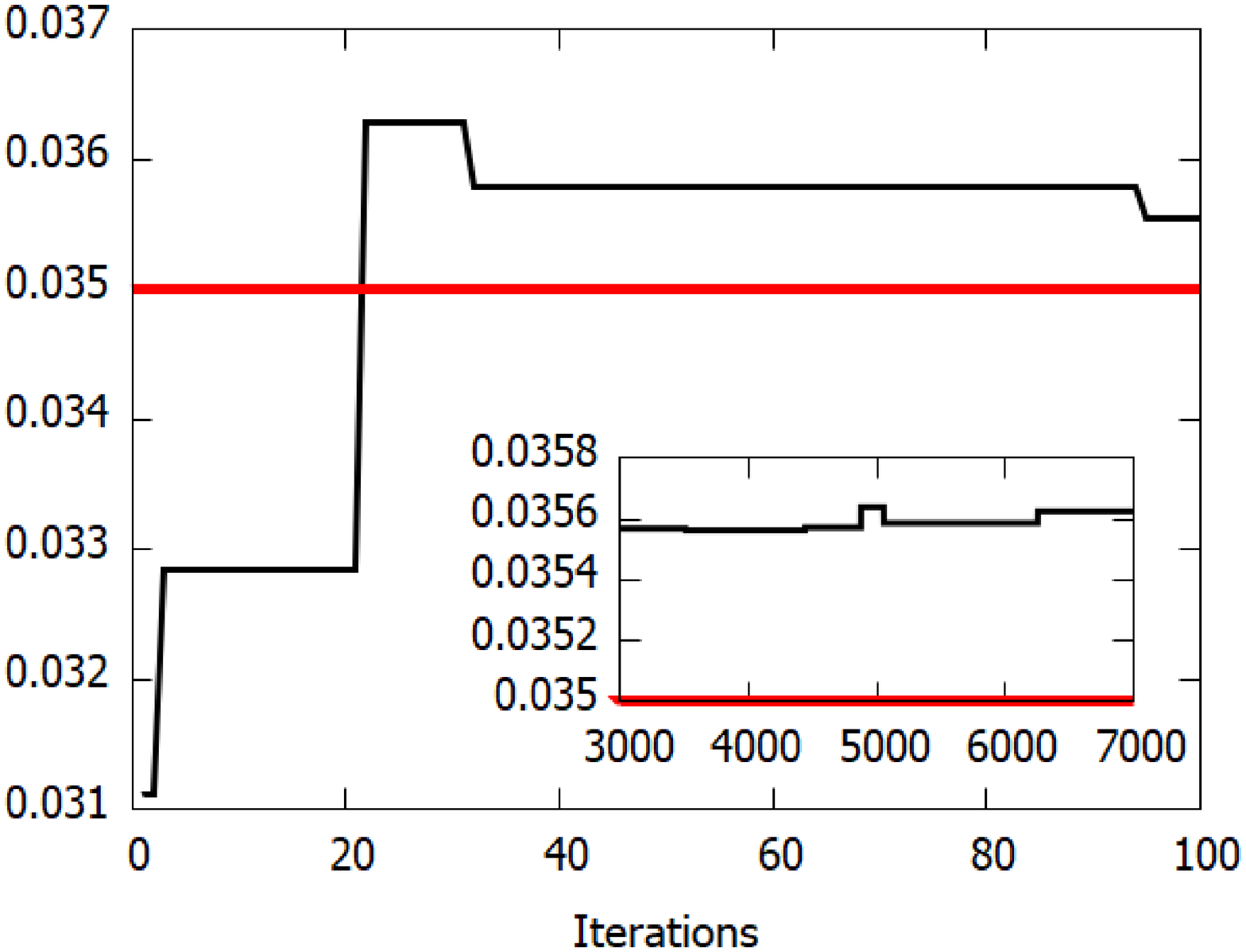}} e) Frequency of detection of persons with active MDR-TB form $\delta_m.$\\
\end{minipage}
\hfill
\begin{minipage}[h]{0.47\linewidth}
\center{\includegraphics[width=1\linewidth]{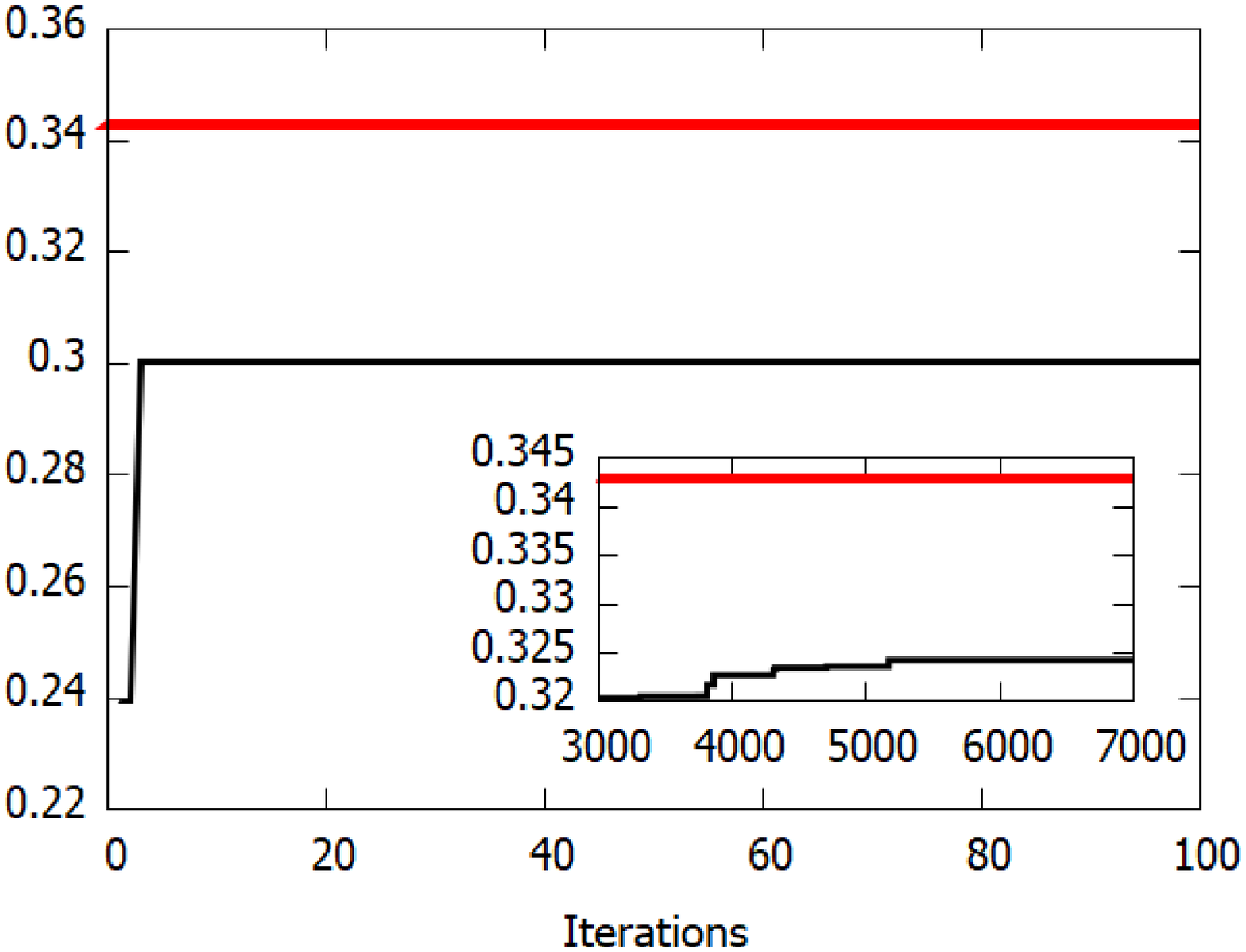}} \\ f) Combination of parameters $\rho \chi$.
\end{minipage}
\vfill

\end{figure}

\begin{table}[!ht]
\caption{Quantitative characteristics of the reconstruction of the parameter vector $q\in\mathbb{R}^7$}\label{tab_parameters}
\begin{tabular}{|l|l|p{2.2cm}|p{2.2cm}|p{2cm}|}
\hline
\multicolumn{2}{|c|}{Parameters} & \multicolumn{3}{c|}{Genetic algorithm} \\ \hline
Symbol & Exact value & Initial approximation & Obtained solution & Common relative error \\ \hline
$\varepsilon$ & 0.129 & 0.16830592 & 0.12789563 & \\
$k$ & 0.821 & 0.67787293 & 0.89007997 &\\
$\nu$ & 0.075 & 0.04805257 & 0.10452100 & \\
$\delta$ & 0.72 & 0.65096273 & 0.63715624 & 0.15236504\\
$\delta_m$ & 0.035 & 0.03111197 & 0.03562611 &\\
$\rho \chi$ & 0.343 & 0.23956309 & 0.32431383 &\\
$o$ & 0.6 & 0.45884955 & 0.43733699 &\\
\hline
\end{tabular}
\end{table}

\section{3D digital Earth system for modeling and prediction of the tuberculosis epidemic}

Tuberculosis is an infectious disease that is widespread in the world. According to the World Health Organization (WHO), every fourth inhabitant of the Earth is a potential carrier of this disease. Although the incidence of tuberculosis in the Republic of Kazakhstan over the past ten years has decreased by 2.4 times and amounted to 52.2 cases per 100 thousand of the population last year, the mortality rate has decreased by almost six times: last year, three cases per 100 thousand population, WHO reports that the Russian Federation, Republic of Kazakhstan and People's Republic of China are in the eighth place in the list of countries with a high incidence of multidrug-resistant tuberculosis (MDR-TB) \cite{12}. 

Tuberculosis is accompanied by quantitative and qualitative changes in the nature of this disease in various regions. To develop an action plan for identifying and treating patients in a particular region, it is necessary to make a forecast of the spread of the epidemic in the region. Mathematical modeling, namely the development of a specific mathematical model that describes the spread of infection, is one of the most effective methods for predicting the spread of the epidemic. In works \cite{13,14} various mathematical models of the spread of tuberculosis, their analysis and a review of modern literature are presented.

One way to build a refined mathematical model is to identify the parameters (coefficients and initial conditions) of existing mathematical models based on the law of mass balance in a closed system, according to some statistics on the number of infected and / or treated individuals at some points in time. To build a short-term forecast map of the epidemiological situation in the region on the basis of an updated model, it is necessary to create a platform that would allow to build maps with good accuracy, manage them and introduce new mathematical models (depending on the region).

To get the high quality prediction map the GIS "Digital Earth'' is developed at the Institute of Computational Mathematics and Mathematical Geophysics SB RAS together with GeoSystem LLC. A feature of the program is a user interface that makes it easy to perform processing, search and three-dimensional visualization of spatial data. The graphical shell of the program is written in C$\#$ and built as a GIS-type mapping system, and it provides easy and efficient manipulation of maps, 3D-models and data. The main options for working with the database are available in the Tools menu and are implemented using the Data Viewer tool, which allows to search, select, sort according to the specified parameters by opening the necessary database.

We will demonstrate the capabilities of GIS "Digital Earth'' for analysis of the epidemiological situations in regions of Russian Federation and Republic of Kazakhstan.

\subsection{Endemic regions of Russian Federation}
The official estimate of the incidence of tuberculosis in Russian Federation is about 82 to 83 per 100,000 population. Russian Federation is a vast country, and the incidence of tuberculosis differs as much as tenfold among geographic regions. It is higher in the eastern portion of the country, with the highest rates in the Russian Far East adjacent to Mongolia, China, and Japan.

International experts estimate that about 50,000 people in Russia have MDR TB. Between 40 and 70 percent of newly detected tuberculosis cases occur in socially vulnerable groups, including the homeless, the unemployed, migrants, and people with drug and alcohol dependencies. The incidence of tuberculosis among the unemployed is 750 per 100,000 unemployed people, compared with 45 per 100,000 employed people. Th most endemic regions in 2015 are Sverdlosk, Novosibirsk, Kemerovo and Far East of Russian Federation. In paper~\cite{1} the prediction of tuberculosis epidemiology in Siberian district (Novosibirsk, Omsk, etc.) is demonstrated based on refinement parameters of mathematical model~(\ref{Krug}) by means of measurements~(\ref{thry}) using combination of stochastic and deterministic approaches. Using proposed approaches with real statistics the tuberculosis maps of Russian Federation are constructed with the help of GIS "Digital Earth''.

A useful function implemented in the GIS is the ability to perform various types of sampling, with which the user can specify the territory. The Select Area item is located on the toolbar, and it allows you to specify the area on the map for selecting from the data list. 

\begin{figure}[!ht]
\center{\includegraphics[width=1\linewidth]{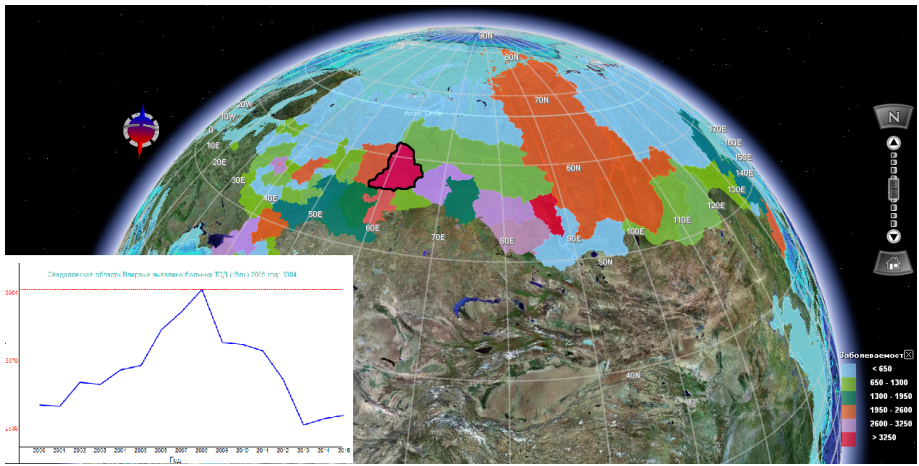}}
\caption{The map of cases of pulmonary tuberculosis registered in 2005. The most endemic regions (red domains with more than 3250 cases per 100000 people) are Sverdlovsk and Kemerovo regions.}
\label{3D-Sverdlovsk_2005_new}
\end{figure}

Figure~\ref{3D-Sverdlovsk_2005_new} demonstrates the map of new diagnosis of tuberculosis in Russian Federation in 2005. The most endemic regions (red domains with more than 3250 cases per 100000 people) are Sverdlovsk (black boundary) and Kemerovo regions. The graph of rate of new diagnosis of tuberculosis in Sverdlovsk region depends on years from 2000 to 2015 and is shown on Figure~\ref{3D-Sverdlovsk_2005_new}.

\begin{figure}[!ht]
\center{\includegraphics[width=1\linewidth]{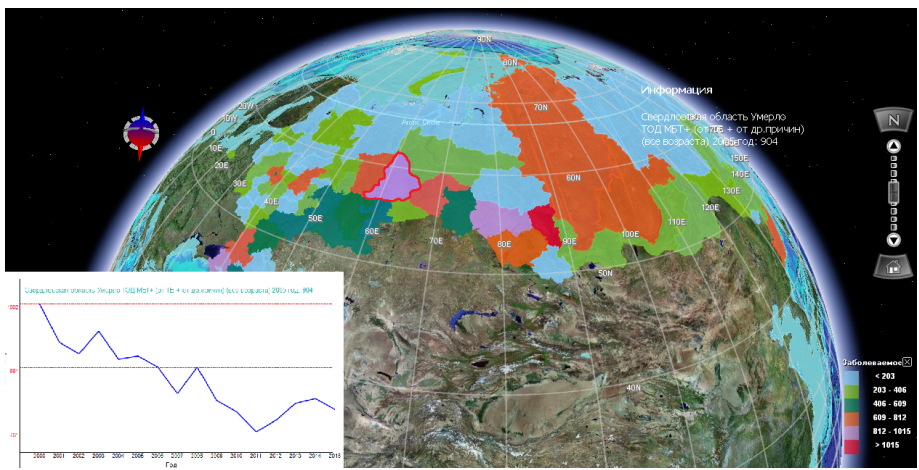}}
\caption{The map of mortality rate of tuberculosis in Russian Federation in 2005. The highest rate (red domain with more than 1015 cases per 100000 people) is in Kemerovo region and high rates (purple domains with 812-1015 cases per 100000 people) are in Sverdlovsk and Novosibirsk regions.}
\label{3D-Sverdlovsk_2005_death}
\end{figure}

Figure~\ref{3D-Sverdlovsk_2005_death} demonstrates the map of mortality rate of tuberculosis in Russian Federation in 2005. The highest rate (red domain with more than 1015 cases per 100000 people) is in Kemerovo region and high rates (purple domains with 812-1015 cases per 100000 people) are in Sverdlovsk (red boundary) and Novosibirsk regions. The graph of rate mortality in Sverdlovsk region depends on years from 2000 to 2015 and is shown on Figure~\ref{3D-Sverdlovsk_2005_death}. 

\subsection{Endemic regions of Republic of Kazakhstan}
Nowadays geographic information technologies in the Republic of Kazakhstan are actively developing. Many organizations are developing network technologies that allow them to accumulate, analyze, and update data without reference to a specific location. Kazakhstan GIS is used in various sectors of the economy and government: land cadastre, geology, production of hydrocarbons, oil and gas transportation, public safety, urban planning, forestry, government, ecology, navigation, etc. A large number of GIS projects are being implemented on the basis of ESRI software, and the ArcGIS platform has become the GIS standard in a number of government departments and large companies. For example, the Academy of Preventive Medicine of Kazakhstan has experience in mapping health (creating electronic maps of the capabilities of the healthcare system) using geographic information systems (GIS) in various regions of Kazakhstan. This approach allows to implement important tools in regional health care, better analyzes the spatial relationships among various factors affecting the health of the population living in this region, allows you to determine the need for additional resources and identify problems and barriers that affect the implementation of health system programs.

\begin{figure}[!ht]
\center{\includegraphics[width=1\linewidth]{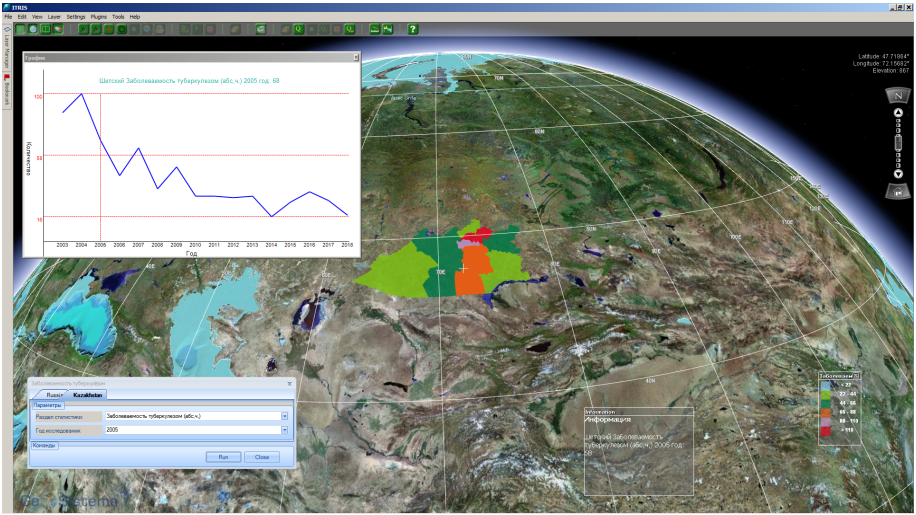}}
\caption{The map of incidence of tuberculosis in the Shetsky district of the Karaganda region in 2005 and a graph of the incidence from 2003 to 2018.}
\label{The incidence of tuberculosis}
\end{figure}

\begin{figure}[!ht]
\center{\includegraphics[width=1\linewidth]{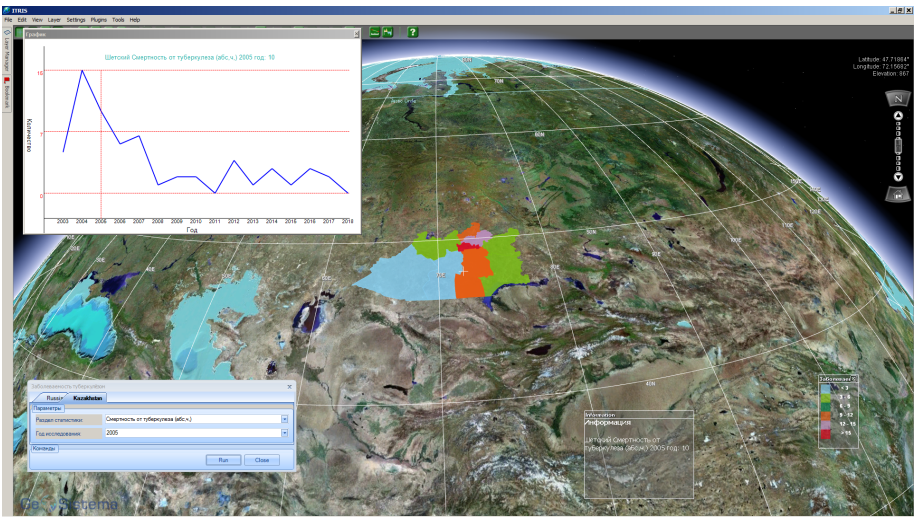}}
\caption{The map of mortality rate of tuberculosis in the Shetsky district of the Karaganda region in 2005 and a graph of rate changes from 2003 to 2018.}
\label{Mortality}
\end{figure}

Using GIS "Digital Earth'' the tuberculosis incidence database for Karaganda region of Republic of Kazakhstan consists of maps divided into regions and is presented on Figures ~\ref{The incidence of tuberculosis}-\ref{Mortality}.

Figures~\ref{3D-Sverdlovsk_2005_new} and~\ref{The incidence of tuberculosis} show the endemic regions in Russian Federation and Republic of Kazakhstan with different prediction in the end.

Figures~\ref{3D-Sverdlovsk_2005_death} and~\ref{Mortality} show the mortality rate of tuberculosis (all cases) decreases in endemic regions in Russian Federation and Republic of Kazakhstan.

\section {Conclusion}
The long-term dynamics of the incidence of tuberculosis (determining the trend, frequency) in regions of Russian Federation and Republic of Kazakhstan in the context of areas characterized by different climatic and socio-economic conditions is studied. Zoning of the territory at various levels of the administrative-territorial division in terms of the incidence of tuberculosis is of great importance both for improving the control strategy and for solving a number of theoretical issues of a biological and environmental nature. Zoning of the most endemic territories in Russian Federation and Republic of Kazakhstan in terms of the incidence of tuberculosis has not yet been carried out. At the same time, such a study is important, because with the widespread spread of tuberculosis in the Karaganda region, the epidemiological significance of different regions is not the same. For more effective anti-TB measures, a deeper epidemiological analysis of the incidence is required.

Timely accumulation of comprehensive information with the introduction of computer processing of data in the sanitary-epidemiological service system will increase the level of information support for epidemiological surveillance. This goal is achieved by using databases as part of geographic information systems. Currently, geoinformation technologies are moving from the stage when they served only as an instrument for integrating data from various sources and as a means of quickly building a variety of maps to spatial geographical analysis and modeling systems and further to full-scale decision support systems.

The problem of refining the mathematical model (namely, its parameters) of the spread of tuberculosis in particularly endemic areas according to statistics from the Karaganda region from 2000 to 2014 is numerically solved. It is shown that the model allows one to characterize the data with good accuracy and gives a short-term forecast of the epidemic prevalence. The obtained data and the mathematical forecast were depicted on the maps of Russian Federation and the Karaganda region of Republic of Kazakhstan using a three-dimensional visualization system.

The future plans about China consist in developing fast numerical method with high-precision for the new mathematical model.

Mathematical modeling is especially important when studying epidemics in large regions including several neighboring countries. On the one hand, the increasing flows of movements leads to coordination of the work of specialists from Shanghai Cooperation Organization~\cite{15} and BRICS~\cite{16, 17, 18}. On the other hand, it leads to visualization of large amounts of data, including transport flows (automobile, railway and aviation), the features of disease spread and various predictive scenarios depending on the adoption of certain measures (vaccination, treatment, quarantine, closing of certain regions and/or transport lines).

\end{document}